%% file: striebel_lattice2011.tex
\documentclass{PoS}
\usepackage{amsmath, amsthm}
\usepackage{float}
\restylefloat{figure}

\newcommand{\onecol}[2]{
        \begin{minipage}[t]{#1}{#2\vfill} \end{minipage}
        } 

\title{Accuracy of Symmetric Partitioned Runge-Kutta Methods for Differential Equations on Lie-Groups}

\ShortTitle{Accuracy of Symmetric Partitioned Runge-Kutta Methods}

\author{
   \hfill
   \onecol{3.5cm}{\vspace{-2.5em}\it
      WUB / 11-23\\
      BUW-IMACM 11-19
   }
   \vspace{1cm}
}

\author{\speaker{Michael Striebel}, Michael G\"unther, Francesco Knechtli and Mich\`ele Wandelt\\
        Bergische Universit\"{a}t Wuppertal, %Lehrstuhl f\"{u}r 
Faculty of Mathematics and Natural Science, Germany
\thanks{This work was supported by the
Deutsche Forschungsgemeinschaft through the Collaborative Research Centre SFB-TR 55 \textquotedblleft Hadron
physics from Lattice QCD \textquotedblright. 
Furthermore, Michael Striebel acknowledges the Marie Curie Initial Training Network STRONGnet 
\textquotedblleft Strong Interaction Supercomputing Training Network\textquotedblright for travel support.} 
        E-mail: \email{ \{striebel, guenther, wandelt\}@math.uni-wuppertal.de}, \email{knechtli@physik.uni-wuppertal.de}}

\abstract{
Computer simulations in QCD are based on the discretization of the theory on a Euclidean lattice. 
To compute the mean value of an observable, usually the Hybrid Monte Carlo method is applied. 
Here equations of motion, derived from an Hamiltonian, have to be solved numerically. 
Commonly the Leapfrog (Stoermer-Verlet) method or splitting methods with multiple timescales \`a la Sexton-Weingarten are used to integrate the dynamical system, defined on a Lie group. \\
Here we formulate time-reversible higher order integrators based on implicit partitioned Runge-Kutta schemes and show that they allow for larger step-sizes than the Leapfrog method. 
Since these methods are based on an infinite series of exponential functions, we concentrate on the truncation of this series with respect to the global error and accuracy.  
Finally, we see that the global error of a SPRK scheme is always even such that a convergence order of one is gained for methods with odd convergence order.

}

\FullConference{ The XXIX International Symposium on Lattice Field Theory - Lattice 2011\\
July 10-16, 2011\\
Squaw Valley, Lake Tahoe, California}

\input{commands}

\begin{document}

%%%%%%%%%%%%%%%%%%%%%%%%%%%%%%%%%%%%
\section{Introduction and Motivation}
\vspace*{-1ex}
In the molecular dynamics step of the Hybrid Monte Carlo method \cite{duane_kennedy_etall:hmc}, Hamiltonian equations of motions have to be solved.
These equations form coupled systems of matrix differential equations of the form
\begin{subequations}
\label{eq:hamilton_nu}
\begin{align}
\dot{\link}_\nu &= \phantom{-}\frac{\partial H\left(\Link,\Moment\right)}{\partial \moment_\nu} 
	=  %f_i \bigl(Y,\Psi\bigr) = 
\moment_\nu \cdot \link_\nu,\label{eq:hamilton_link_nu}\\
\dot{\moment}_\nu &= - \frac{\partial H\left(\Link,\Moment\right)}{\partial \link_\nu} = g(\link_\nu), \qquad \text{for $\nu=1,\dotsc, n$}.\label{eq:hamilton_momentum_nu}
\end{align}
\end{subequations}
In this notation, $\link_\nu$ is an element of a matrix Lie group $\group$ and $\moment_\nu$ an element of its associated Lie algebra $\algebra$. 
Thus, $\Link$ can be imagined as a vector of $n$ matrix Lie group elements $\link_1, \link_2, \ldots, \link_n$, and $\Moment$ as a vector of $n$ Lie algebra elements $\moment_1, \moment_2, \ldots, \moment_n$.
%%%%%%%%%%%%%%%%%%
In pure lattice gauge theory, the element $\link_\nu$ can be seen as the link $U_{x,\mu}$ between the lattice sites $x$ and $x+a\hat{\mu}$. 
Thus, $\moment_\mu$ is its associated momentum $P_{x,\mu}$ times the complex $i$. 
In this context, the vectors of matrices $\Link$ and $\Moment$ are the whole configurations of the links and its momenta. \\
%%%%%%%%%%%%%%%%%%%%%%
The equations of motion have to be solved in a Lie group, respectively in a Lie algebra with a time-reversible and area-preserving scheme. 
In a recent paper \cite{SPRK}, we have investigated the potentialities of higher order partitioned Runge-Kutta schemes for solving the equations of motions such that the desired properties are met.
We found out that symmetric partitioned Runge-Kutta methods based on the Magnus and Munthe-Kaas approach can be time-reversible. 
So far, area-preservation is not fulfilled and must be corrected in the acceptance step.
Furthermore, the global error of this scheme is always even
%%%%%%%%%%%%%%%%%%%%%%%%%%%
and investigated in detail in this paper.  
In doing so, we start with a short derivation of symmetric partitioned Runge-Kutta schemes based on the ideas of Magnus and Munthe-Kaas.  
Afterwards, we focus on the global error and accuracy of the method and show some numerical results.
%%%%%%%%%%%%%%%%%%%%%%%%%%%
\vspace*{-1ex}
\section{Numerical Integration}
\vspace*{-1ex}
The differential equations \eqref{eq:hamilton_nu} become an initial value problem (IVP) by prescribed initial values: $\link_\nu(0):= \link_{\nu,0}$ and $\moment_\nu(0):=\moment_{\nu,0}$ for $\nu=1,\dotsc,n$. 
Thereby, the initial values $\link_\nu(0)$ have to be in the Lie group and the elements $\moment_\nu(0)$ in the Lie algebra. 
Considering the structure of 
equation \eqref{eq:hamilton_nu}, \eqref{eq:hamilton_link_nu} is a differential equation in a Lie group such that it has to be solved with a numerical scheme that guarantees a solution in the Lie group as described in paragraph \ref{section:de}
%%%%%%%%%%%%%%%%%%%%%%%%%%%
For the second equation \eqref{eq:hamilton_momentum_nu}, no special treatment has to be applied. It is an equation in the Lie algebra $\algebra$, which is a linear space. 
Thus, this equation can be solved with any time-reversible and area-preserving numerical scheme.
%%%%%%%%%%%%%%%%%%%%%%%%%%%
For convenience, we leave out the index $\nu$ from now on. 
This means we investigate just one coupled differential equation for a special but arbitrary index $\nu$: 
$\dot{\link} = \moment \cdot \link$ and %\label{eq:hamilton_link} &
$\dot{\moment} = g(\link)$. %
% \begin{subequations}
% %\label{eq:hamilton}
% \begin{align*}
% % \dot{Y}_k &= \Psi_k(Y(t))\cdot Y_k(t) = f_i(Y_k(t),\Psi_k(Y(t)),\label{eq:problem_link}\\
% % \dot{\Psi}_k(Y) &= g_k(Y_1(t),\dotsc, Y_n(t)) \quad\text{for $k=1,\dotsc,n$}\label{eq:problem_momentum},
% \dot{\link} &= \moment \cdot \link,&%\label{eq:hamilton_link} &
% \dot{\moment} &= g(\link).%\label{eq:hamilton_momentum}
% \end{align*}
% \end{subequations}
The results can then be extended straightforward to the whole vectors $\Link$ and $\Moment$.
\vspace*{-1ex}
\subsection{Differential Equations in Lie Groups}\label{section:de}
\vspace*{-1ex}
Concerning equation \eqref{eq:hamilton_link_nu}, we follow the ideas of Magnus and Munthe-Kaas. 
Magnus \cite{magnus:exp_solution} stated that the differential equation \eqref{eq:hamilton_link_nu} in the Lie group can be replaced through a differential equation in the Lie algebra. 
This new differential equation can be solved directly due to the linearity of the Lie algebra. 
The strategy is as follows: 
Identify $\link(t)$ with $\exp(\alink(t))$ such that the variable changes from $\link$ to $\alink$. $\alink$ is the solution of the differential equation
   \begin{align}
\label{eq:magnus}
 \dot \alink = d \exp_{\alink}^{-1}(\moment),
\end{align}
with $\alink(t)\in\algebra$ and initial value $\alink(0):=0$. 
The derivative of the inverse exponential map \eqref{eq:magnus} is given by an infinite series as
\begin{align*}%\label{eq:dexpinv}
 d \exp_{\alink}^{-1} \left(\moment\right) =\sum_{k \ge 0} \frac{B_k}{k!} ad_{\alink}^k \left(\moment\right).
\end{align*}
In this series, the variables $B_k$ are the $k$-th Bernoulli numbers and the adjoint operator $ad_{\alink}^k$ is a mapping in the Lie algebra $\algebra$ given by
% \begin{align*}%\label{adjoint_operator}
%   \ad_{\alink}(\moment) &:=[\alink,\moment]=\alink\moment - \moment\alink
% \end{align*} 
  $\ad_{\alink}(\moment) :=[\alink,\moment]=\alink\moment - \moment\alink$.
It follows the conventions
  $\ad_{\alink}^0(\moment) = \moment$ and
  $\ad_{\alink}^k(\moment) = [\alink, \ad_{\alink}^{k-1}(\moment)]$.
This means, $\alink$ is the solution of the differential equation 
\begin{align*}%\label{Alink_infinite}
 \dot \alink &= \sum_{k \ge 0} \frac{B_k}{k!} ad_{\alink}^k \left(\moment\right).
\end{align*}
Knowing $\alink$, the solution $\link$ of \eqref{eq:hamilton_link_nu} can be attained via 
% \begin{align*}
%  \link=\exp(\alink)\link_{0}.
% \end{align*}
$\link=\exp(\alink)\link_{0}$.
In total, we record that the initial value problem \eqref{eq:hamilton_nu} is equivalent to 
% \begin{subequations}
% \label{eq:mtran}
% \begin{align}
% \dot{\alink} &= \sum_{k=0}^{\infty} \frac{B_k}{k!} \ad_{\alink}^k(\moment), \label{eq:mtran_alink}\\
% \dot{\moment} &= g(\link)
% \quad\text{with}\;\;\link=\exp(\alink)\link_{0},\label{eq:mtran_moment}
% \end{align}
% \end{subequations}
\begin{align}\label{eq:mtran}
\dot{\alink} &= \sum_{k=0}^{\infty} \frac{B_k}{k!} \ad_{\alink}^k(\moment), 
&
\dot{\moment} &= g(\link)
\quad\text{with}\;\;\link=\exp(\alink)\link_{0} %\label{eq:mtran_moment}
\end{align}
%\end{subequations}
and $\link(0):=\link_{0} \in \group$, $\moment(0):=\moment_{0} \in \algebra$ and $\alink(0):=0\in\algebra$.
%%%%%%%%%%%%%%%%%%%%%%%%%%%
This transformed problem can now be solved directly by a Runge-Kutta method without destroying the Lie group structure: 
As the Lie algebra $\algebra$ is a vector space, the analytic solution $(\alink(t), \moment(t))$ as well as its approximation $(\alink_1, \moment_1)$ attained by a numerical integration scheme both are elements of the Lie algebra $\algebra$. 
Furthermore, as for any $a\in \algebra$ the matrix exponential $\exp(a)$ is in the associated matrix Lie group $\group$, also $\link$ is in $\group$.
%%%%%%%%%%%%%%%%%%%%%%%%%%%
\vspace*{-1ex}
\subsection{Symmetric Partitioned Runge-Kutta schemes}
\vspace*{-1ex}
The problem in solving \eqref{eq:mtran} is that $\dot \alink$ is given as infinite series which has to be suitably truncated after $q+1$ terms. 
This means, the truncation index $q$ of $\dot \alink$
has to be chosen properly
such that a numerical integration scheme meets a prescribed convergence order $p$. 
Thereby, the convergence order of a numerical integration method is $p$ if the deviation between the exact solution and its numerical approximation after one step is of order $p+1$ in a suitable norm.
Here, the idea of Munthe-Kaas comes into play. 
He states in \cite{Munthe-Kaas99highorder} that the truncation index $q$ of $\dot \alink$
has to be chosen as a value larger than the desired convergence order $p$ minus one.
Consequently, for a Runge-Kutta scheme of convergence order $p$, $\dot \alink$ is set as a function depending on the truncation $q \geq p-2$ of the aforementioned infinite series, \ie 
\begin{align}\label{Alink_finite}
 \dot \alink &= \sum_{k = 0}^{q} \frac{B_k}{k!} ad_{\alink}^k \left(\moment\right) =: f_q(\alink,\moment).
\end{align}
%%%%%%%%%%%%%%%%%%%%%%%%%%%
All in all, the exact solution of \eqref{eq:mtran} is approximated through an integration scheme of order $p$ of the truncated model
%\begin{subequations}
\begin{align}\label{eq:mtrantrunc}
\dot{\widehat{\alink}} &= \sum_{k=0}^{q =p-2} \frac{B_k}{k!} \ad_{\widehat{\alink}}^k(\widehat{\moment}), %\label{eq:mtrantrunc_alink}\\
&
\dot{\widehat{\moment}} &= g(\widehat{\link})
\quad\text{with}\;\;\widehat{\link}=\exp(\widehat{\alink})\widehat{\link}_{0},%\label{eq:mtrantrunc_moment}
\end{align}
%\end{subequations}
$\widehat{\link}(0):={\link}_{0} \in \group$, $\,\widehat{\moment}(0):={\moment}_{0} \in \algebra$ and $\widehat{\alink}(0):=0\in\algebra$.
%%%%%%%%%%%%%%%%%%%%%%%%%%%
Thereby, this model can be solved with higher order time-reversible symmetric partitioned Runge-Kutta (SPRK) schemes derived in \cite{SPRK} as follows:
Compute the approximations
% \begin{subequations}
% \label{eq:sprk_ode_lie}
% \begin{align}
% \alink_1 &=  h\sum_{i=1}^sb_i K_i, & \moment_1 &= \moment_0 + h\sum_{i=1}^s \widehat{b}_i L_i,
% \intertext{with increments $K_i$ and $L_i$ for $i=1,\dotsc,s$ defined by}
% K_i &= f_q\left(\bar{\alink}_i, \bar{\moment}_i\right), & L_i&=g\left(\bar{\link}_i\right),
% \label{eq:sprk_ode_lie_increments}
% \end{align}
% {In the course of this, the internal stages are defined as}
% \begin{align}
% \bar{\alink}_i &= h\sum_{j=1}^s \alpha_{ij}K_j, &
% \bar{\moment}_i &= \moment_0+h\sum_{j=1}^s \widehat{\alpha}_{ij}L_j,\\
% \bar{\link}_i &=\exp\left(\bar{\slink}_i\right)\exp\left(\tfrac{1}{2}\alink_1\right)\link_0, &\bar{\slink}_{i} &= h\sum_{j=1}^s \gamma_{ij} K_j,
% \label{eq:sprk_ode_lie_stages}
% \end{align}
% \end{subequations}
\begin{align}\label{eq:sprk_ode_lie}
\alink_1 &=  h\sum_{i=1}^sb_i K_i, & \moment_1 &= \moment_0 + h\sum_{i=1}^s \widehat{b}_i L_i,
\end{align}
with increments $K_i = f_q(\bar{\alink}_i, \bar{\moment}_i)$ and $L_i=g\left(\bar{\link}_i\right)$ for $i=1,\dotsc,s$.
In the course of this, the internal stages are defined as
% \begin{align}
% \bar{\alink}_i &= h\sum_{j=1}^s \alpha_{ij}K_j, &
% \bar{\moment}_i &= \moment_0+h\sum_{j=1}^s \widehat{\alpha}_{ij}L_j,\\
% \bar{\link}_i &=\exp\left(\bar{\slink}_i\right)\exp\left(\tfrac{1}{2}\alink_1\right)\link_0, &\bar{\slink}_{i} &= h\sum_{j=1}^s \gamma_{ij} K_j,
% \label{eq:sprk_ode_lie_stages}
% \end{align}
\begin{align*}
\bar{\alink}_i &= h\sum_{j=1}^s \alpha_{ij}K_j, &
\bar{\moment}_i &= \moment_0+h\sum_{j=1}^s \widehat{\alpha}_{ij}L_j,&
\bar{\link}_i &=\exp\left(\bar{\slink}_i\right)\exp\left(\tfrac{1}{2}\alink_1\right)\link_0, &\bar{\slink}_{i} &= h\sum_{j=1}^s \gamma_{ij} K_j.
%\label{eq:sprk_ode_lie_stages}
\end{align*}
At the end, the solution $\link_1$ is attained via $\link_1=\exp(\alink_1)\link_0$.
In this scheme, the coefficients $b_i, \widehat{b}_i, \alpha_{ij}, \widehat{\alpha}_{ij}$ and $\gamma_{ij}$ for $i,j = 1,\ldots,s$ can be determined to guarantee time-reversibility (and symmetry). 
Their values for convergence order $p=3$ can be found in \cite{SPRK}.
\vspace*{-1ex}
\section{Global Error and Accuracy of the SPRK Method}
\vspace*{-1ex}
For the local error, the solution of the integration method after one step has to be compared with the exact solution $\link(t_0+h), \moment(t_0+h)$ of the differential equations \eqref{eq:hamilton_nu}.
The SPRK method \eqref{eq:sprk_ode_lie} is of convergence order $p$ if
% As explained by Munthe-Kaas, the SPRK method \eqref{eq:sprk_ode_lie} is of local order $p$ with respect to the Lie group differential equation \eqref{eq:hamilton_nu} if 
\begin{equation}\label{eq:convergence_order_sprk}
\|\link(t_0+h) -\link_1\| = \mathcal{O}(h^{p+1}) \quad \text{and}\quad \|\moment(t_0+h) -\moment_1\| = \mathcal{O}(h^{p+1})
\end{equation}
holds. 
Since the approximation to $\link_1$ is computed from evaluating the matrix exponential (we assume here that we can evaluate this exactly), which is Lipschitz on every closed interval, it suffices to demand 
\begin{equation*}%\label{eq:convergence_order_sprk_algebra}
\|\alink(t_0+h) -\alink_1\| = \mathcal{O}(h^{p+1}) \quad \text{and}\quad \|\moment(t_0+h) -\moment_1\| = \mathcal{O}(h^{p+1})
\end{equation*}
with exact solution $\alink(t_0+h), \moment(t_0+h)$. 
%%%%%%%%%%%%%%%%%%%%%%%%%%%
%\subsection{Accuracy}\label{sec:munthekass}
According to Munthe-Kaas, the approximations $\alink_1$ and $\moment_1$ of the exact solution of the suitably truncated problem \eqref{eq:mtrantrunc} 
can also be interpreted as approximations to the original problem \eqref{eq:mtran}.
With the same argument, we can even formulate a stronger statement on the local accuracy \eqref{eq:convergence_order_sprk}. 
As the method is symmetric, theorem~3.2 in \cite[II.3]{hairer_lubich_wanner:gni} applies, which states that the maximal convergence order $p$ of a symmetric method is even, which means that the local error is always odd.
Hence, the SPRK method developed as a method of an odd convergence order $p$ is of order $p+1$. 
%In Sec.~\ref{section:ordercondition} we have constructed conditions for the method's coefficients such that the local order is $p=3$. Hence, due to the symmetry the order is increased by one and in fact the local errors reported in \eqref{eq:convergence_order_sprk} are of order $p=4$, \ie, $\mathcal{O}(h^5)$.
%}
%%%%%%%%%%%%%%%%%%%%%%%%%%%
The global error of a numerical integrated scheme is computed as the sum of the local errors. 
This means, for the computation of a trajectory with length $\tau$, an integration method with fixed step size $h$ is applied $N = \tau / h$ times.
Hence, the global error is of the order local error minus one. Thus, the SPRK method of convergence order $p$ has at least a global error of order $p$. 
Again, in case of an odd $p$, the global error is of order $p+1$.\\ 
%%%%%%%%%%%%%%%%%%%%%%%%%%%
The accuracy depends on the truncation $k=q=p-2$ of the series given in \eqref{eq:mtran} with $p$ being the convergence order of the Runge-Kutta method that is used to solve the problem numerically. 
We recall the basic steps of the proof given in \cite{hairer_lubich_wanner:gni} for a deeper understanding of the choice of the truncation parameter $q$. 
For this purpose, we restrict to an uncoupled Lie algebra problem
\begin{align}
	\dot{\alink} = F_\infty(\alink) := \sum_{k=0}^{\infty} \frac{B_k}{k!}\ad_\alink^k(A) , \quad \text{with}\;\; \alink(0)=0\in \algebra, \label{eq:laproblem}
\end{align}
which arises from the Magnus approach mentioned in paragraph \ref{section:de}.
% \begin{align}\label{eq:lgproblem}
% 	\dot{\link}&=A(\link)\link, \quad\text{with}\;\; \link(0)=\link_0\in \group,
% \intertext{where $A:\group \rightarrow \algebra$.
 The truncation of the series in \eqref{eq:laproblem} at $k=q$ yields the truncated problem
\begin{align}
	\dot{\widehat{\alink}} &= F_q(\widehat{\alink}), \quad \text{with}\;\; \widehat{\alink}(0)=0, \label{eq:laproblem_trunc}
\end{align}
{such that}
\begin{align*}
F_{\infty}(\alink(t)) - F_q(\alink(t)) = \sum_{k=q+1}^{\infty} \frac{B_k}{k!}\ad_{\alink(t)}^k(A).
\end{align*}
For sufficiently smooth $A$ we recognize
\begin{align*}
	\ad_{\alink(t)}^k (A) = \mathcal{O}(t^{k+1}),
\end{align*}
which is due to the nested structure of the $\ad$-Operator \cite{hairer_lubich_wanner:gni}. Hence, we have
% \begin{equation}
% 	\begin{split}
% 		F_q(\widehat{\alink}(t)) &=  F_\infty(\widehat{\alink}(t))  + C(t), \\
% 	&\text{with}\;\; C(t) = c_1 t^{q+2}+c_2 t^{q+3}+\dotsb
% 	\end{split}
% \end{equation} 
\begin{equation*}
		F_q(\widehat{\alink}(t)) =  F_\infty(\widehat{\alink}(t))  + C(t)
% 	&\text{with}\;\; C(t) = c_1 t^{q+2}+c_2 t^{q+3}+\dotsb
% 	\end{split}
\end{equation*} 
with $C(t) = c_1 t^{q+2}+c_2 t^{q+3}+\dotsb$ and constant values $c_1, c_2, \dotsc \in\algebra$. 
%%%%%%%%%%%%%%%%%%%%%%%%%%%
For fixed $h>0$ and $t\in[0,h]$ the exact solutions $\alink(t)$ of \eqref{eq:laproblem} and $\widehat{\alink}(t)$ of \eqref{eq:laproblem_trunc} satisfy
\begin{equation}\label{eq:exact}
\begin{split}
	\|\alink(t)-\widehat{\alink}(t)\| &= \|\alink(0)+\int_0^t F_\infty(\alink(\tau))\,\text{d}\tau - \Bigl(\widehat\alink(0)+\int_0^t F_q(\widehat\alink(\tau))\,\text{d}\tau\Bigr) \|\\
	&= \left\Vert\int_0^t F_\infty(\alink(\tau))\,\text{d}\tau - \left(\int_0^t F_\infty(\widehat\alink(\tau))+C(\tau)\,\text{d}\tau \right)\right\Vert\\
	&\le \int_0^t \|F_\infty(\alink(\tau))-F_\infty(\widehat\alink(\tau))\|\,\text{d}\tau + \int_0^t\|C(\tau)\|\,\text{d}\tau.
\end{split}
\end{equation}
The function $F_\infty$ is Lipschitz continuous on every closed interval for sufficiently smooth $A$. 
We assume that for an interval where both $\alink(t)$ and $\widehat\alink(t)$ reside in for $t\in[0,h]$, the Lipschitz constant is $L_\infty\in \R$, \ie, 
\begin{align*}
\|F_\infty(\alink(t))-F_\infty(\widehat\alink(t))\| \le L_\infty\|\alink(t)-\widehat\alink(t)\|.
\end{align*}
Furthermore, for $t\in[0,h]$ we see that
\begin{align}\label{eq:C}
	\|C(t)\| %&\le \|c_1\|t^{q+2}+\|c_2\|t^{q+3}+\dotsb\\
	&\le \|c_1\|h^{q+2}+\|c_2\|h^{q+3}+\ldots \; := \bar{c}(h) \in \R.
\end{align}
Hence from \eqref{eq:exact} it follows that 
\begin{align*}
	\|\alink(t)-\widehat{\alink}(t)\| \le \bar{c}(h)\cdot t + L_\infty \int_0^t\|\alink(\tau)- \widehat\alink(\tau)\|\,\text{d}\tau,
\end{align*}
such that the requirements of the "Gronwall lemma" \cite{hairer_norsett_wanner:solving1} are satisfied by which
\begin{align*}
	\|\alink(t)-\widehat\alink(t)\| %&\le \frac{\bar{c}(h)}{L_\infty}\left(e^{L_\infty t}-1\right)\\
	&\le \frac{\bar{c}(h)}{L_\infty}\left(e^{L_\infty h}-1\right)\\
	%&= \frac{1}{L_\infty}\left(\|c_1\|h^{q+2}+\|c_2\|h^{q+3}+\dotsb\right)\\&\quad\quad\cdot \left(\left(1+L_\infty h + \frac{1}{2!}(L_\infty h)^2 + \dotsb\right)-1\right)\\
	&= \frac{1}{L_\infty}\left(\|c_1\|h^{q+2}+\|c_2\|h^{q+3}+\dotsb\right) \cdot \Bigl(\bigl(1+L_\infty h + \frac{1}{2!}(L_\infty h)^2 + \dotsb\bigr)-1\Bigr)\\
	&=\left(\|c_1\|h^{q+2}+\|c_2\|h^{q+3}+\dotsb\right)\cdot\Bigl(h+\frac{1}{2}L_\infty h^2 +\frac{1}{3!}L_\infty^2h^3+\dotsb\Bigr).
\end{align*}
Thus, the difference between the exact solutions of the full problem \eqref{eq:laproblem} and the truncated problem \eqref{eq:laproblem_trunc} is 
\begin{align}
	\|\alink(h)-\widehat\alink(h)\|= \mathcal{O}(h^{q+3}) \label{eq:difference_exact}
\end{align}
after one time step $h$.
Applying a one step method of convergence order $p$ on the truncated problem \eqref{eq:laproblem_trunc}  means to calculate an approximation $\widehat{\alink}_1$ to the exact value $\widehat\alink(h)$ such that
\begin{align}
	\|\widehat{\alink}(h)-\widehat\alink_1\| = \mathcal{O}(h^{p+1}). \label{eq:difference_approximation}
\end{align}
Finally, we interpret $\widehat\alink_1$ as an approximation to the exact solution of the original problem \eqref{eq:laproblem}. 
The quality of this approximation is determined by the deviation \eqref{eq:difference_exact} introduced by the modeling and the discretization error \eqref{eq:difference_approximation}:
% \begin{equation}\label{eq:difference}
% 	\begin{split}
% 		\|\alink(h)-\widehat\alink_1\| &\le \| \alink(h)-\widehat\alink(h)\| + \| \widehat\alink(h) - \widehat\alink_1\|\\
% 		&= \mathcal{O}(h^{q+3}) + \mathcal{O}(h^{p+1}).
% 	\end{split}
% \end{equation} 
\begin{equation*}%\label{eq:difference}
		\|\alink(h)-\widehat\alink_1\| \le \| \alink(h)-\widehat\alink(h)\| + \| \widehat\alink(h) - \widehat\alink_1\|
		= \mathcal{O}(h^{q+3}) + \mathcal{O}(h^{p+1}).
\end{equation*} 
This clearly indicates that $\widehat\alink_1$ is a numerical approximation to $\alink(h)$ of convergence order $p$, \ie, 
$$\|\alink(h)-\widehat{\alink}_1\|=\mathcal{O}(h^{p+1}) \qquad \text{if}\;\; q+3\ge p+1, \quad\text{\ie,}\;\; q\ge p-2.$$
%%%%%%%%%%%%%%%%%%%%%%%%%%%
%%%%%%%%%%%%%%%%%%%%%%%%%%%%%%%%%%
\vspace*{-6ex}
\section{Numerical Tests}
\vspace*{-1ex}
% \begin{figure}[ht]
% \centering
%  \includegraphics[width=7cm]{../Lattice2011/absDeltaH_beta2_8x8_sfb} 
%  \hspace{0.1cm}
%  %\includegraphics[width=7cm]{} 
%  \caption{Solutions of the SHO; explicit Euler method with step size $h=0.1$, initial value $(p_{0},~q_{0}) =(0,1)$;
% implicit Euler method with step size $h=0.1$, initial value $(p_{0},~q_{0}) =(0,1)$.}
% \label{fig:1}
% \end{figure}
We consider a pure lattice gauge theory in SU(2,$\C$) with Wilson action and compare the SPRK method described in \eqref{eq:sprk_ode_lie} with the Leapfrog method. 
For this purpose, we investigate a symmetric partitioned Runge-Kutta scheme of convergence order $p=4$ which contains the truncated function $\dot \alink = f_q(\alink,\moment)$ given in \eqref{Alink_finite}. 
Because of the symmetry, the method has an even convergence order such that the choice $p=3, q=1$ already leads to a local error of order 5. 
This means, we use the equation
\begin{equation}\label{eq:f1}
 \dot \alink = f_1(\alink,\moment) = \moment -\frac12 [\alink, \moment]
\end{equation}
according to \eqref{Alink_finite} 
and perform simulations on a 2-dimensional lattice with lattice size $L=T=32$.
\begin{figure}[b]
  \begin{centering}\hspace*{0.5cm}
    \includegraphics[width=0.37\textwidth]{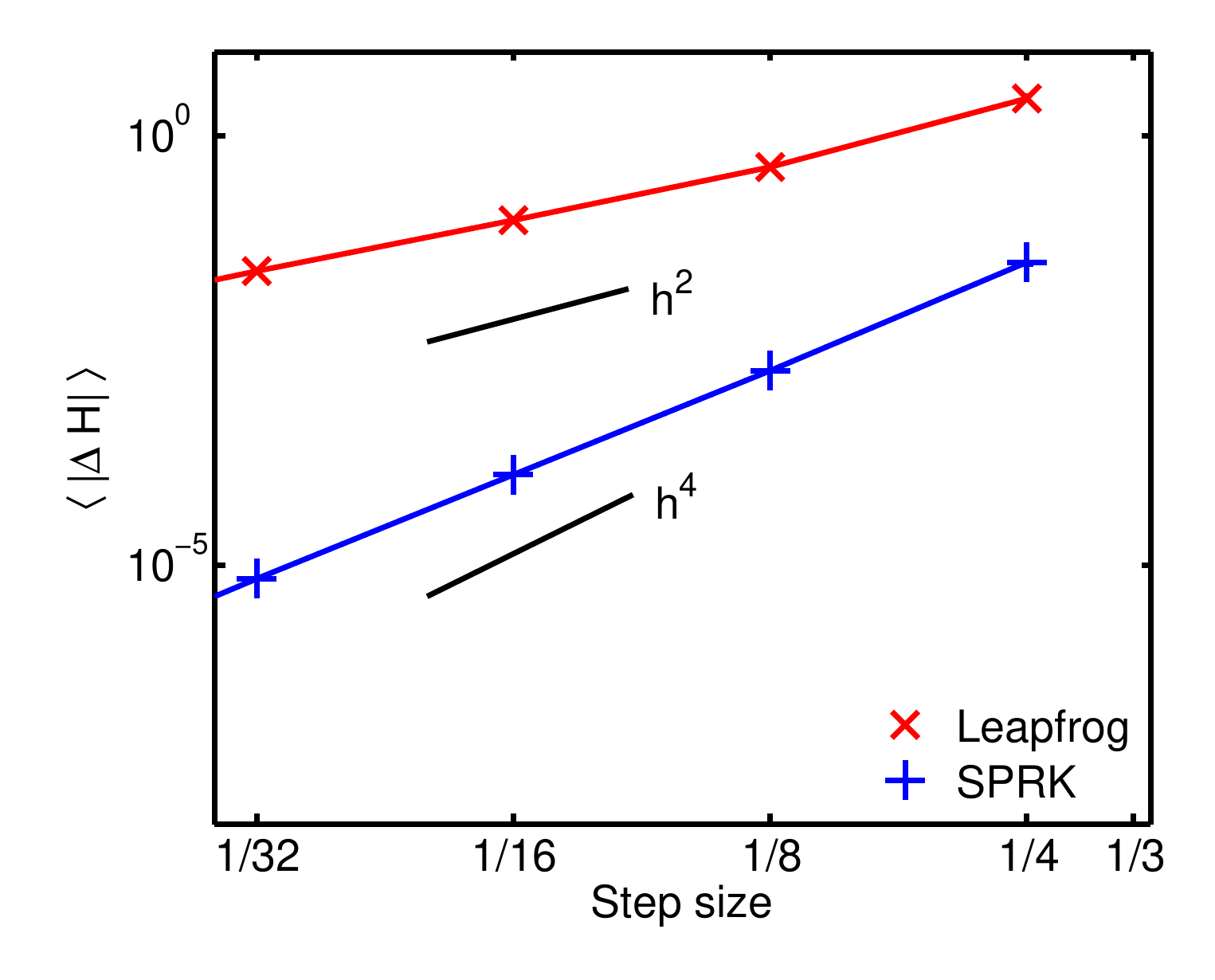}%\hspace*{1cm}
     \hspace*{1cm}
    \includegraphics[width=0.37\textwidth]{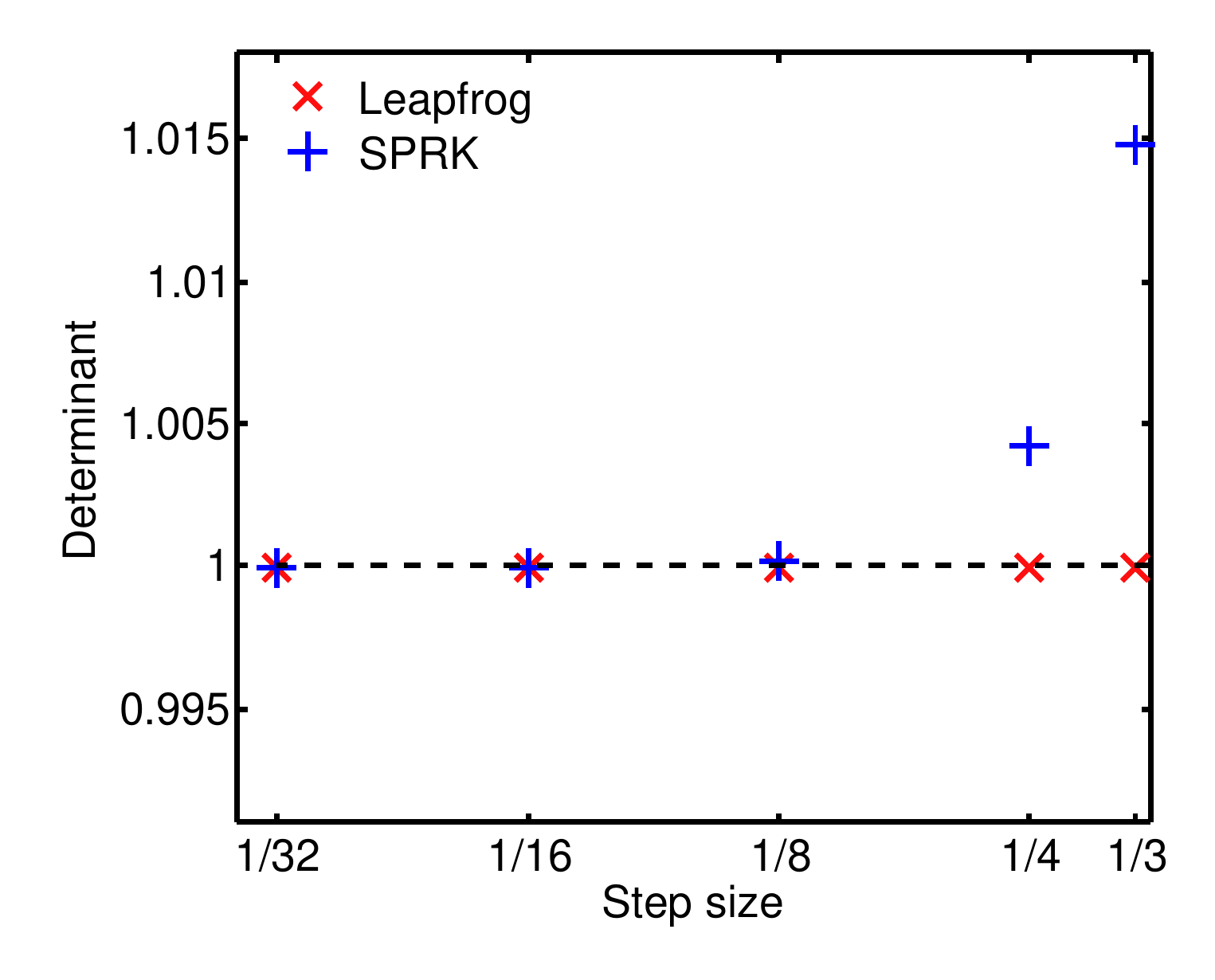}%{det_8x8_err1e05}%symp_8x8_eps1e-5}
    \end{centering}
\caption{Left: Convergence order.
  Right: Area-preservation.}
\label{fig:1}
\end{figure}
There are 2 results shown in figure~\ref{fig:1}: On the left side, the convergence order of the different methods can be seen. For this purpose, we consider the energy change $\Delta H$ of two successive configurations
after a whole trajectory of length 1  and take the mean of 5000 configurations. The statistical errors are so small that they are not visible in the plot.
Since the energy change $\Delta H$ deviates from zero just because of the numerical errors of the integration scheme, the violation of the energy preservation gives the global error. 
We see that for a given $|\Delta H|$ the SPRK allows for larger step sizes.
On the right side of figure~\ref{fig:1}, we see the violation of the area-preservation in dependence of the step sizes chosen in the numerical methods. 
Area-preservation (up to roundoff errors) is given if the determinant of $\partial(\alink_1, \moment_1) / \partial (\alink_0, \moment_0)$ has exactly the value 1. 
Here, the determinant is numerically approximated by first order difference quotients.

%%%%%%%%%%%%%%%%%%%%%%%%%%
%%%%%%%%%%%%%%%%%%%%%%%%%%
\vspace*{-1ex}
\section{Conclusion}
\vspace*{-1ex}
We investigated the accuracy of the time-reversible symmetric partitioned Runge-Kutta scheme \eqref{eq:sprk_ode_lie}. 
The order of accuracy consists of two components: On the one hand, the convergence order depends of course on the order $p$ of the method itself.
As the method is symmetric, the local error is always odd, \ie, the scheme has a local error of order $p+1$ for an even convergence order $p$.
%%%%%%%%%%%%%%%%%%%%%%%%%%%
On the other hand, the SPRK scheme contains one truncated series \eqref{Alink_finite}. The truncation index $q$ has to be larger than or equal to $p-2$ to meet the prescribed convergence order.
%%%%%%%%%%%%%%%%%%%%%%%%%%%
All in all, choosing an SPRK method with an even local error should be preferred since a convergence order of one is gained by the symmetry.
We performed simulations for an SPRK scheme of convergence order 4 and see that the global error given in the numerical results has order 4 as theoretically expected. 
In the development of the SPRK method, area-preservation has not been considered. 
Thus, it is not surprising, that area-preservation is not met applying this scheme. 
This property has to be investigated in future work.  
% 
% For efficiency, the order $p$ of the method should be preferably high, as this allows for large step-sizes $h$ to satisfy prescribed error tolerances. \\
%%%%%%%%%%%%%%%%%%%%%%%%%%%
%%%%%%%%%%%%%%%%%%%%%%%%%%%
%%%%%%%%%%%%%%%%%%%%%%%%%%%%%%%%%%

%%%%%%%%%%%%%%%%%%%%%%%%%%%
%%%%%%%%%%%%%%%%%%%%%%%%%%%%%%%%%%
%\bibliographystyle{plain}
%\bibliography{sLPRK.bib}

%%%%%%%%%%%%%%%%%%%%%%%%%%%
\end{document}

%% file: commands.tex
\newcommand{\Link}{[U]}
\newcommand{\link}{U}
\newcommand{\Moment}{[A]}
\newcommand{\moment}{A}
\newcommand{\alink}{\Omega}

\newcommand{\slink}{X}

\newcommand{\ad}{\text{ad}}

\newcommand{\group}{G}
\newcommand{\algebra}{\mathfrak{g}}
\newcommand{\ie}{i.~e.}

\newcommand{\C}{\mathbb{C}}
\newcommand{\R}{\mathbb{R}}

%% file: striebel_lattice2011.bbl
\vspace*{-1ex}
\begin{thebibliography}{99}
%%%%%%%%%%%%%%%%%%%%%%%%%%%
\vspace*{-1ex}
\bibitem{duane_kennedy_etall:hmc}
  S. Duane, A. D. Kennedy, B. J. Pendleton and D. Roweth,
  \emph{Hybrid {M}onte {C}arlo},
  \emph{Physics Letters B}, 216--222, 195, 1987.
% 	Date-Modified = {2011-04-29 16:32:44 +0200},
% 	Journal = {Physics Letters B},
% 	Pages = {216--222},
% 	Title = {Hybrid {M}onte {C}arlo},
% 	Volume = 195,
% 	Year = 1987}
%%%%%%%%%%%%%%%%%%%%%%%%%%%
\bibitem{SPRK}
  M.~Wandelt, M.~G{\"u}nther, F.~Knechtli, M.~Striebel,
  \emph{Symmetric {P}artitioned {R}unge-{K}utta {M}ethods},
  2011,
  {\tt arXiv:1109.3030 [hep-lat]}.
\bibitem{magnus:exp_solution}
	Wilhelm Magnus,
	\emph{On the exponential solution of differential equations for a linear operator},
	\emph{Communications on Pure and Applied Mathematics},
	%Wiley Subscription Services, Inc., A Wiley Company,
	4, 
	649--673,
	{\tt doi:10.1002/cpa.3160070404}.
%%%%%%%%%%%%%%%%%%%%%%%%%%%
\bibitem{Munthe-Kaas99highorder}
	Hans Munthe-Kaas,
	\emph{High order {R}unge-{K}utta methods on manifolds},
	\emph{Appl. Numer. Math},
	29, 
	115--127,
	1999.
%%%%%%%%%%%%%%%%%%%%%%%%%%%
% \bibitem{hairer2000} 
% E. Hairer, 
% \emph{Symmetric projection methods for differential equations on manifolds}, 
% \emph{BIT}  {\bf 40} (2000) 726.
% %, pp. 726-734.
% %Abstract. Projection methods are a standard approach for the numerical solution of differential equations on manifolds. It is known that geometric properties (such as symplecticity or reversibility) are usually destroyed by such a discretization, even when the basic method is symplectic or symmetric. In this article, we introduce a new kind of projection methods, which allows us to recover the time-reversibility, an important property for long-time integrations. 
%%%%%%%%%%%%%%%%%%%%%%%%%%%
%%%%%%%%%%%%%%%%%%%%%%%%%%%
%%%%%%%%%%%%%%%%%%%%%%%%%%%
\bibitem{hairer_norsett_wanner:solving1}
	Hairer, E and N{\o}rsett, S. P. and Wanner, G.,
	\emph{Solving {O}rdinary {D}ifferential {E}quations I -- {N}onstiff {P}roblems},
	  \emph{Springer},
	  second revised,
	  2000.
%%%%%%%%%%%%%%%%%%%%%%%%%%%
\bibitem{hairer_lubich_wanner:gni}
E.Hairer, C.Lubich and G.Wanner, 
\emph{Geometric Numerical Integration Structure-Preserving 
          Algorithms for Ordinary Differential Equations}, 
Springer Ser.\ Comput.\ Math.\ {\bf 31}, 2nd ed.,
Springer, 2006.
%%%%%%%%%%%%%%%%%%%%%%%%%%%
\bibitem{Kamleh:2011dc}
W.~Kamleh and M.~Peardon,
\emph{Polynomial Filtered HMC -- an algorithm for lattice QCD with dynamical
quarks},
{\tt arXiv:1106.5625 [hep-lat]}.
%%%%%%%%%%%%%%%%%%%%%%%%%%%
\bibitem{Clark:2011ir}
M.~A.~Clark, B.~Jo{\'o}, A.~D.~Kennedy, P.~J.~Silva,\\
\emph{Improving dynamical lattice QCD simulations through integrator tuning
  using Poisson brackets and a force-gradient integrator},
{\tt arXiv:1108.1828 [hep-lat]}. 
%%%%%%%%%%%%%%%%%%%%%%%%%%%
% 	Date-Added = {2011-03-05 16:29:04 +0100},
% 	Date-Modified = {2011-03-05 16:29:04 +0100},
% 	Edition = {second revised},
% 	Publisher = {Springer},
% 	Title = {Solving {O}rdinary {D}ifferential {E}quations I -- Nonstiff {P}roblems},
% 	Year = {2000}}.
%%%%%%%%%%%%%%%%%%%%%%%%%%%
% \bibitem{hairer_wanner:solving2}
%   Hairer, E and Wanner, G.
%   \emph{Solving {O}rdinary {D}ifferential {E}quations II -- {S}tiff and {D}ifferential-{A}lgebraic {P}roblems}
%   \emph{Springer}
%   second revised,
%   1996.
% 	Date-Added = {2011-03-05 16:29:04 +0100},
% 	Date-Modified = {2011-03-05 16:29:04 +0100},
% 	Edition = {second revised},
% 	Publisher = {Springer},
% 	Title = {Solving {O}rdinary {D}ifferential {E}quations II -- {S}tiff and {D}ifferential-{A}lgebraic {P}roblems},
% 	Year = {1996}}.
\end{thebibliography}
